\documentclass[10pt,twocolumn,a4paper]{article}

\usepackage[top=0.8in,bottom=1in,left=0.75in,right=0.75in,columnsep=0.25in]{geometry}

\usepackage[utf8]{inputenc}
\usepackage[T1]{fontenc}
\usepackage{amsmath,amssymb,amsfonts}
\usepackage{times}
\usepackage{microtype}
\usepackage{graphicx}
\usepackage{xcolor}
\usepackage{textcomp}
\usepackage{multirow}
\usepackage{booktabs}
\usepackage{listings}
\usepackage{tcolorbox}
\usepackage{enumitem}
\setlist{itemsep=2pt,topsep=3pt,parsep=0pt,leftmargin=1.6em}

\usepackage[font=small,labelfont=bf,labelsep=period]{caption}
\usepackage[compact]{titlesec}
\titleformat{\section}{\centering\normalfont\normalsize\scshape}{\thesection.}{0.6em}{}
\titleformat{\subsection}{\normalfont\normalsize\itshape}{\thesubsection}{0.6em}{}
\titleformat{\subsubsection}{\normalfont\small\itshape}{\thesubsubsection}{0.6em}{}
\titlespacing*{\section}{0pt}{1.4ex plus .3ex minus .2ex}{0.8ex plus .1ex}
\titlespacing*{\subsection}{0pt}{1.2ex plus .3ex minus .2ex}{0.6ex plus .1ex}
\titlespacing*{\subsubsection}{0pt}{1.0ex plus .3ex minus .2ex}{0.5ex plus .1ex}

\usepackage{cite}

\usepackage{authblk}

\renewenvironment{abstract}{\small\noindent\textbf{\textit{Abstract}---}\bfseries}{\par\vspace{4pt}}

\usepackage[hidelinks,
    pdftitle={AERIAL: Mission-Aware LLM Selection for Language-Driven UAV Control in Flying Networks},
    pdfauthor={Diana Nunes, Ricardo Amorim, Pedro Ribeiro, Andr\'e Coelho, Rui Campos}]{hyperref}

\begin{document}

\title{AERIAL: Mission-Aware LLM Selection for Language-Driven UAV Control in Flying Networks%
\thanks{This work is financed by National Funds through the FCT --
Funda\c{c}\~ao para a Ci\^encia e a Tecnologia, I.P. (Portuguese Foundation for
Science and Technology) within the project FALCON, with reference
2023.15645.PEX (\url{https://doi.org/10.54499/2023.15645.PEX}).}}

\author{Diana Nunes}
\author{Ricardo Amorim}
\author{Pedro Ribeiro}
\author{Andr\'e Coelho}
\author{Rui Campos}
\affil{INESC TEC and Faculdade de Engenharia, Universidade do Porto, Porto, Portugal\\
\normalfont\ttfamily\small
\{diana.m.nunes, pedro.m.ribeiro, andre.f.coelho, rui.l.campos\}@inesctec.pt, ricardoaraujoamorim@protonmail.com}
\date{}

\maketitle

\begin{abstract}
Unmanned Aerial Vehicles (UAVs) in Flying Networks (FNs) increasingly require mechanisms to translate high-level natural language objectives into executable missions. Although Large Language Models (LLMs) have demonstrated the feasibility of language-driven UAV control, existing approaches typically rely on static model configurations and overlook mission-aware performance variability. This paper presents AERIAL, a benchmark-driven framework for adaptive LLM selection in hybrid edge-cloud UAV control. AERIAL separates systematic model benchmarking from mission execution, deriving empirical scenario-model mappings and applying them at runtime for mission-aware selection. Evaluation across 720 automated simulation runs spanning navigation, spatial reasoning, and algorithm invocation revealed significant performance variability depending on scenario-model pairing. A case study integrating AERIAL with a FNs placement algorithm further validated the framework's ability to interface with domain-specific optimization modules through natural language. These results validate the effectiveness of AERIAL's adaptive model selection in improving reliability and operational robustness for autonomous UAV control in FNs.
\end{abstract}

\noindent{\small\textbf{\textit{Keywords}---}Flying Networks, Unmanned Aerial Vehicles, Large Language Models, Adaptive Model Selection, Edge-Cloud Computing, Autonomous Control.}

\section{Introduction}
Unmanned Aerial Vehicles (UAVs) have evolved from isolated aerial platforms into cooperative networked systems capable of forming Flying Networks (FNs) that provide communications coverage, sensing, and computing services. In these networks, UAVs execute mission objectives while maintaining connectivity, adapting to traffic conditions, and operating under strict energy and safety constraints \cite{Ribeiro2024}.

Conventional FN control approaches typically rely on predefined mission scripts or manual reconfiguration. While effective in static settings, these approaches limit
scalability and responsiveness in dynamic environments. As FN deployments become more heterogeneous and mission requirements more variable, there is a growing
need for interaction mechanisms that allow operators to specify objectives at a higher level of abstraction. This motivates the development of frameworks capable of
interpreting natural language instructions and translating them into executable UAV control logic that respects both mission goals and networking constraints.

Recent work has explored the use of Large Language Models (LLMs) for natural language-driven UAV control. The FLUC framework~\cite{fluc2025} demonstrated that locally deployed LLMs can generate executable mission code interfaced with standard autopilot systems. Although FLUC established the feasibility of offline LLM-based control, it relied on a single, manually selected model. Experimental results presented in~\cite{fluc2025} showed that model performance varied significantly across mission types, particularly in missions requiring structured spatial reasoning. In addition,
FLUC did not consider cloud-assisted reasoning or systematic model benchmarking, limiting its adaptability across different deployment scenarios. These results indicate that LLM-based UAV control in FNs should not rely on a fixed model configuration; model selection should be informed by empirical performance across different mission categories.

To address this limitation, this paper introduces AERIAL, a benchmark-driven framework for mission-aware LLM selection in hybrid edge-cloud FN environments. AERIAL adopts a two-phase architecture that separates model evaluation from mission execution. During the benchmarking phase, multiple LLMs are evaluated through automated simulation-based UAV missions, allowing quantitative characterization of their performance across heterogeneous mission categories. In the execution phase, incoming natural language commands are classified and routed to the model that achieved the best empirical performance for the corresponding mission type. In addition, AERIAL supports hybrid edge-cloud operation. Locally deployed open-source models ensure offline autonomy, while cloud-based LLMs can be invoked when connectivity permits and additional reasoning capability is required. This design supports operation under different connectivity conditions, allowing local execution when cloud access is unavailable or not required, as well as cloud-based reasoning when connectivity is available.

The main contributions of this paper are as follows:
\begin{itemize}
\item \textbf{The AERIAL framework}, a benchmark-driven two-phase architecture that separates model benchmarking from mission execution, incorporating an adaptive LLM selection mechanism based on empirically learned performance profiles and a hybrid edge-cloud execution model that ensures offline autonomy while enabling cloud-based reasoning when connectivity is available;

\item An \textbf{extensive experimental evaluation} comprising 720 automated, simulation-based test runs across diverse UAV mission scenarios, demonstrating significant mission-dependent performance differences among LLMs and validating the effectiveness of the proposed adaptive model selection strategy.

\item \textbf{A case study} integrating AERIAL with a state of the art UAV placement algorithm, validating the framework's ability to interface with domain-specific optimization modules through natural language.
\end{itemize}

The central question this paper addresses is whether mission-aware LLM selection produces measurably better outcomes than static single-model assignment for UAV control, and whether model-mission affinities are strong enough to justify the added complexity of an adaptive selection mechanism. To answer this, AERIAL provides a systematic benchmarking methodology that quantifies model performance across mission categories and drives selection policies directly from empirical data.

The remainder of this paper is organized as follows. Section~\ref{RW} reviews related work on LLM-based UAV control, code generation, adaptive model selection, and UAV positioning in FNs. Section~\ref{AC} describes the AERIAL framework, including the two-phase design and hybrid edge-cloud integration. Section~\ref{IM} details the AERIAL framework implementation and system components. Section~\ref{EM} outlines the experimental methodology and evaluation scenarios, and Section~\ref{ER} presents the corresponding results. Section~\ref{CS} reports a case study integrating AERIAL with the SUPPLY UAV placement algorithm. Section~\ref{D} discusses implications and limitations. Finally, Section~\ref{C} concludes the paper and outlines directions for future work.

\section{Related Work}\label{RW}
Recent advances in LLMs have raised a growing interest in natural language-driven UAV control, LLM-based code generation, and intelligent networking for FNs. This section reviews related work across three directions: i) LLM-based UAV control; ii) LLM-driven code generation for networking systems; and iii) adaptive model selection and hybrid edge-cloud execution. Our goal is to highlight limitations related to fixed model usage, limited benchmarking practices, and the absence of mission-aware model selection mechanisms.

\subsection{LLM-Based UAV Control}

Several recent works have explored the integration of LLMs into UAV systems. Tian et al.~\cite{Tian_2025} provide a systematic review of LLM-UAV integration and propose an agentic framework for autonomous perception and reasoning in low-altitude mobility. Their work organizes applications across perception, planning, and control, offering a conceptual foundation. However, it is fundamentally architectural and does not include empirical validation in high-fidelity simulators or flight missions.

Voice-driven control has also received attention. Choutri et al.~\cite{Choutri2025} present a bilingual voice-based UAV control system combining offline speech recognition with LLM-based command interpretation. Their architecture achieves low end-to-end latency and demonstrates practical usability. Nevertheless, semantic reasoning is performed by cloud-based LLMs, which limits deployment in connectivity-constrained environments.

Other related works have explored LLM-based mission generation in simulation environments. Vemprala et al.~\cite{Vemprala2024} demonstrated the feasibility of using ChatGPT \cite{chatgpt2023} for robotic control and code synthesis in AirSim \cite{airsim2017}. Tazir et al.~\cite{Tazir2023} integrated ChatGPT with PX4 \cite{px4} and Gazebo \cite{gazebo} for basic UAV command interpretation. These works confirm the practicality of natural language-driven control, but rely exclusively on cloud APIs and do not explore systematic benchmarking or local model execution.

Javaid et al.~\cite{Javaid2024_2} provide a survey of potential LLM applications in UAV autonomy, including perception and decision support. However, their contribution remains conceptual and does not present an implemented system or comparative evaluation across models.

More structured architectures have recently emerged. Wang et al.~\cite{Wang2025Chat} propose a dual-agent design that separates task planning from execution, showing improved success rates in complex tasks. LLM-Land~\cite{LLMLand2025} demonstrates context-aware UAV landing using lightweight local models and retrieval-augmented generation, showing that small LLMs can support safety-critical tasks. The NeLV framework \cite{yuan2025nextgenerationllmuavnatural} introduces a taxonomy of automation levels for LLM-based UAV systems, ranging from parser-based assistance to autopilot-level integration. While these works advance the architectural discussion, they do not incorporate systematic benchmarking across heterogeneous mission categories.

In parallel, Chagas et al.~\cite{Chagas2025} review AI-based UAV deconfliction methods, highlighting the dominance of reinforcement learning in tactical conflict resolution. Their analysis indicates that the integration of LLM-based reasoning with established collision avoidance techniques remains unexplored.

The FLUC framework~\cite{fluc2025} represents one of the first fully local LLM-based UAV control implementations using the Ollama~\cite{ollama} runtime and ArduPilot~\cite{ardupilot}. FLUC demonstrates that open-source models can generate executable mission code for coordinate navigation, semantic geolocation, and obstacle avoidance. However, it relies on a fixed model configuration and does not include systematic performance benchmarking or model selection mechanisms.

\subsection{LLM-Based Code Generation for Networking Systems}

Beyond UAV control, LLM-based code generation has been studied in networking contexts. Ahmed et al.~\cite{simcode2025} introduced SIMCODE, a benchmark for evaluating LLMs in generating ns-3 ~\cite{ns3} network simulation code from natural language descriptions. Their results show substantial variability in code correctness and reliability across models, reinforcing the importance of systematic benchmarking. Rezazadeh et al.~\cite{Rezazadeh2025} extend this direction with a multi-agent framework that automates the full simulation lifecycle in ns-3 from natural language parsing through script generation, test execution, and result interpretation, demonstrating that LLM-driven pipelines can reduce manual coding effort in complex 5G/6G network environments.

The similarities between network simulation code generation and UAV mission synthesis are direct. Both tasks require translating natural language specifications into executable code within domain-specific APIs, ensuring correctness under structural constraints, and managing safety or performance requirements. These similarities motivate the application of structured benchmarking methodologies to UAV control.

\subsection{Adaptive Model Selection and Hybrid Edge-Cloud Execution}

Although benchmarking of LLMs has received increasing attention \cite{Sun2024_2,Rong2024}, adaptive model selection for autonomous systems remains relatively underexplored. Mon-Williams et al.~\cite{ELLMER2025} demonstrate that LLMs can support delay-tolerant reasoning in embodied robotics using retrieval-augmented generation. However, their approach relies on cloud-hosted models and does not consider model selection strategies across heterogeneous tasks.

Within UAV networking, Sun et al.~\cite{Sun2024} explore LLM-enabled coordination graphs for dynamic networking, while Sun et al.~\cite{Sun2024_2} analyze generative AI applications in UAV networks. These works emphasize coordination and networking intelligence but do not address model selection or hybrid edge-cloud execution policies.

Hao et al.~\cite{Hao2024} investigate hybrid edge-cloud inference, showing that small edge models can collaborate with large cloud-hosted models through dynamic routing strategies. Their findings support the feasibility of splitting inference workloads across distributed resources, providing a technical basis for hybrid deployment architectures.

\begin{table*}[t]
\centering
\small
\setlength{\tabcolsep}{4pt}
\caption{Comparison of related work in LLM-based UAV control and code generation.}
\label{tab:comparison}
\begin{tabular}{|l|c|c|c|c|c|c|c|c|c|}
\hline
\textbf{Feature} & \textbf{Nunes} & \textbf{Tian} & \textbf{Choutri} & \textbf{Vemprala} & \textbf{Tazir} & \textbf{Wang} & \textbf{Cai} & \textbf{Yuan} & \textbf{Ahmed} \\
 & \textbf{et al.} & \textbf{et al.} & \textbf{et al.} & \textbf{et al.} & \textbf{et al.} & \textbf{et al.} & \textbf{et al.} & \textbf{et al.} & \textbf{et al.} \\
 & \textbf{2025} & \textbf{2025} & \textbf{2025} & \textbf{2024} & \textbf{2023} & \textbf{2025} & \textbf{2025} & \textbf{2025} & \textbf{2025} \\
 & \cite{fluc2025} & \cite{Tian_2025} & \cite{Choutri2025} & \cite{Vemprala2024} & \cite{Tazir2023} & \cite{Wang2025Chat} & \cite{LLMLand2025} & \cite{yuan2025nextgenerationllmuavnatural} & \cite{simcode2025} \\
\hline
Adaptive Selection & $\times$ & $\times$ & $\times$ & $\times$ & $\times$ & $\times$ & $\times$ & $\times$ & $\times$ \\
\hline
Automated Benchmarking & $\times$ & $\times$ & $\times$ & $\times$ & $\times$ & $\times$ & $\times$ & $\times$ & \checkmark \\
\hline
Cloud LLM & $\times$ & $\times$ & \checkmark & \checkmark & \checkmark & \checkmark & $\times$ & \checkmark & \checkmark \\
\hline
Hybrid Execution & $\times$ & $\times$ & \checkmark & $\times$ & $\times$ & $\times$ & $\times$ & $\times$ & $\times$ \\
\hline
Local LLM & \checkmark & \checkmark & $\times$ & $\times$ & $\times$ & $\times$ & \checkmark & \checkmark & $\times$ \\
\hline
Performance Learning & $\times$ & $\times$ & $\times$ & $\times$ & $\times$ & $\times$ & $\times$ & $\times$ & $\times$ \\
\hline
UAV Control Focus & \checkmark & \checkmark & \checkmark & \checkmark & \checkmark & \checkmark & \checkmark & \checkmark & $\times$ \\
\hline
\end{tabular}
\end{table*}

\subsection{Summary}
Table~\ref{tab:comparison}  summarizes the key architectural features across the related works. The comparison reveals that existing frameworks focus on either local or cloud-based LLM execution, with most relying on fixed model configurations. While SIMCODE demonstrates the value of systematic benchmarking for network code generation, and several systems achieve effective UAV control through natural language interfaces, to the best of our knowledge, no prior UAV control framework simultaneously combines automated LLM benchmarking, performance-based model learning, adaptive mission-aware model selection, and hybrid edge--cloud execution. This gap motivates the need for new approaches that can systematically evaluate LLMs and dynamically select models based on empirical performance data rather than manual configuration.

\section{AERIAL Framework}\label{AC}

\begin{figure*}[t]
\centering
\includegraphics[width=\textwidth]{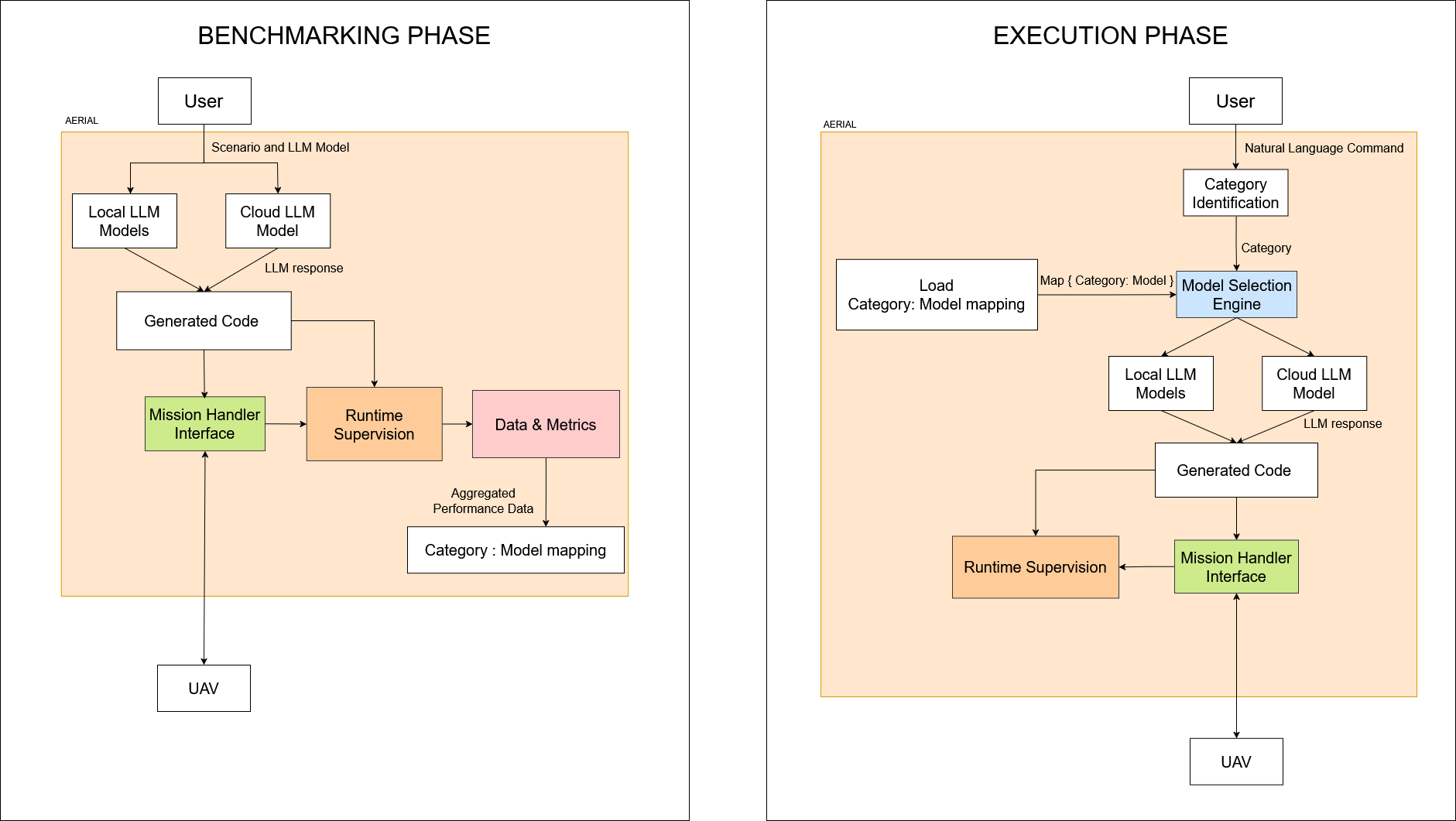}
\caption{AERIAL's two-phase architecture. The Benchmarking Phase (left) performs automated benchmarking and generates scenario-model performance profiles. The Execution Phase (right) uses the resulting mapping to allocate incoming natural language missions to selected models.}
\label{fig:architecture}
\end{figure*}

AERIAL is a benchmark-driven framework for mission-aware LLM selection in hybrid edge-cloud UAV control. The architecture separates empirical model evaluation from operational deployment. As shown in Fig.~\ref{fig:architecture}, AERIAL operates in two phases: a \textbf{Benchmarking Phase}, where LLMs are systematically benchmarked across mission categories, and an \textbf{Execution Phase}, where mission prompts are routed according to the resulting performance profiles.

\subsection{Two-Phase Architecture}

The two-phase design decouples performance profiling from real-time mission execution.

During the \textbf{Benchmarking Phase}, multiple LLMs are evaluated using a predefined set of UAV mission scenarios. For each model-scenario combination, the system generates mission code, executes it in simulation, and validates the outcome through runtime supervision. Performance metrics, including response latency, token usage, and mission success rate, are recorded for each run. The aggregated results are used to construct a scenario-model mapping, which captures the empirically observed mission-aware performance of each model. This mapping serves as the basis for mission-aware selection during operation.

During the \textbf{Execution Phase}, incoming natural-language prompts are first classified into one of the predefined mission categories (coordinates, location, diagonal, waypoints, obstacle, or algorithm). The corresponding model is then selected through deterministic lookup of the precomputed scenario-model mapping. No runtime model comparison is performed, ensuring predictable execution overhead while preserving the benefits of benchmark-driven selection.

\subsection{System Components}

AERIAL comprises four primary modules.

The \textbf{Model Selection Engine} performs lookup-based model selection using the precomputed scenario-model mapping generated during the Benchmarking Phase. It supports both local inference through an embedded runtime and cloud-based inference through provider-agnostic APIs. The modular design allows additional models to be integrated without modifying the overall architecture.

The \textbf{Mission Handler Interface} translates generated high-level commands into autopilot protocol messages and manages communication with the flight controller. This layer encapsulates low-level interaction details and ensures proper command sequencing.

The \textbf{Runtime Supervision Layer} operates in both phases to monitor mission execution using continuous telemetry tracking. It observes UAV position, altitude, and motion state, and determines movement completion using a stability threshold based on consecutive position readings. During the Benchmarking Phase, this layer additionally applies scenario-specific validation criteria to determine mission success (this is elaborated in Section~\ref{AF:Workflow}).

The \textbf{Data and Metrics Manager} stores benchmarking results, execution logs, and the scenario-model mapping. Structured performance records enable periodic re-evaluation when models are updated or new candidates are introduced.

\subsection{Workflow} \label{AF:Workflow}

The workflow follows the two-phase structure shown in Fig.~\ref{fig:architecture}.

\textbf{Benchmarking Phase:} Predefined mission scenarios (e.g. the ones presented in Section~\ref{subsec:scenarios}) are executed for each candidate LLM. The Runtime Supervision Layer validates mission outcomes using scenario-specific criteria, including coordinate proximity, waypoint traversal, obstacle clearance, and trajectory characteristics. Performance metrics, including response time, token usage, prompt count, and success outcome, are recorded and aggregated to construct the scenario-model mapping.

\textbf{Execution Phase:} An operator provides a natural-language mission request. A lightweight classifier assigns the prompt to one of the predefined mission categories. The Model Selection Engine retrieves the corresponding model from the stored mapping. The selected model generates executable mission code, which is passed through the Mission Handler Interface to the autopilot. The Runtime Supervision Layer monitors execution but does not apply automated validation during deployment.

This separation ensures that performance evaluation occurs under controlled benchmarking conditions, while operational execution remains lightweight and deterministic.

\subsection{Key Architectural Features}

AERIAL incorporates several architectural characteristics that address limitations observed in prior LLM-based UAV control frameworks.

First, AERIAL supports hybrid edge-cloud operation, allowing inference to be performed either by locally deployed models or by cloud-based models depending on connectivity conditions. Local models provide full offline autonomy, while cloud inference can be used when available and when additional reasoning capability is desirable. This dual-mode design enables operation under varying network conditions typical of FNs.

Second, model selection is driven by empirical performance data collected during the Benchmarking Phase. AERIAL constructs a scenario-model mapping that captures mission-aware behavior across mission categories. During execution, incoming prompts are classified into predefined categories, and the Model Selection Engine retrieves the corresponding model through deterministic lookup. The mapping can be updated periodically as new models are introduced or performance profiles change.

Third, the automated benchmarking pipeline enables systematic and reproducible benchmarking. Model-scenario combinations are executed programmatically in simulation and mission outcomes are validated using scenario-specific criteria. This process reduces subjective evaluation and produces structured datasets that characterize model performance across heterogeneous missions. The emphasis on structured benchmarking is consistent with methodologies applied in related domains. For example, SIMCODE~\cite{simcode2025} shows that comparative evaluation across multiple LLMs reveals mission-specific performance patterns in network code generation. AERIAL adopts a similar approach for UAV mission synthesis.

Finally, the framework maintains compatibility with standard UAV control stacks through modular separation of classification, model selection, code generation, and mission execution. This structure allows individual components to be updated or extended without modifying the overall architecture.

Together, these architectural features distinguish AERIAL from prior work in LLM-based UAV control. To the best of our knowledge, no previous framework combines local and cloud-based model support, hybrid execution, automated benchmarking, performance learning, adaptive mission-aware model selection, and direct UAV control within a single architecture.

\section{Implementation}\label{IM}

AERIAL integrates automated benchmarking, performance tracking, and mission-aware model selection within a unified control framework. This section describes its current implementation, including the software stack, core components, and data management structure.

\subsection{Software Stack}

AERIAL is developed in Python 3 and integrates the following technologies.

\textbf{Ollama Runtime} provides local LLM inference through a REST API endpoint. It supports execution of open-source models with CPU or GPU acceleration and enables fully offline operation without external dependencies.

\textbf{Cloud API Integration} enables access to cloud-hosted LLMs when connectivity permits. The current implementation uses Google's Gemini API via the official Python client (\texttt{google.genai}). The interface is provider-agnostic and supports integration of alternative services (e.g., OpenAI, Anthropic, Azure OpenAI) through adapter layers without modifying core system logic. API credentials are managed through environment variables using \texttt{python-dotenv}.

\textbf{ArduPilot Software-In-The-Loop (SITL)}~\cite{ardupilot_sitl} provides a simulation environment for mission execution without physical hardware. SITL emulates autopilot behavior, telemetry streams, and navigation subsystems, allowing validation of LLM-generated control logic prior to deployment. Because SITL uses the same MAVLink protocol interface as physical hardware, validated mission code can be directly transferred to real UAVs \cite{ArduPilotValidation2018, Koubaa2017}.

\textbf{MAVProxy}~\cite{mavproxy} acts as the communication bridge between AERIAL and ArduPilot, translating Python-based mission commands into MAVLink messages~\cite{mavlink}. Communication is established via UDP.

\textbf{OpenStreetMap}~\cite{openstreetmap} is used for semantic location resolution. In scenarios involving named locations, AERIAL queries the Nominatim API to retrieve geographic coordinates corresponding to natural language references.

\subsection{Core Components}

The implementation consists of the following modular Python scripts that support both the Benchmarking and Execution phases:

\begin{itemize}
    \item \texttt{ardupilot\_\allowbreak ollama.py} implements the primary control loop. It manages user interaction, model selection, code generation, and execution orchestration. It supports both benchmarking mode (via the \texttt{--test} flag) and interactive execution.
    \item \texttt{task\_\allowbreak classifier.py} performs prompt classification using LLaMA 3.2 to assign incoming commands to one of six predefined categories (coordinates, location, diagonal, waypoints, obstacle, algorithm). The resulting category is used for scenario-model lookup.
    \item \texttt{automated\_\allowbreak tests.py} implements the benchmarking pipeline. The \texttt{AutomatedTester} class coordinates model-scenario execution, applies validation functions, and records structured test results.
    \item \texttt{test\_\allowbreak validator.py} defines scenario-specific validation logic within the \texttt{TestValidator} class. It verifies mission outcomes using telemetry-based criteria, including coordinate proximity, waypoint traversal, obstacle avoidance, and trajectory constraints.
    \item \texttt{extract\_\allowbreak metrics.py} aggregates test results from JSON logs and generates structured CSV datasets (\texttt{combined\_\allowbreak models\_\allowbreak metrics.csv}) containing response latency, token counts, success rates, and 95\% confidence intervals.
    \item \texttt{update\_\allowbreak model\_\allowbreak map.py} analyzes aggregated metrics and produces the \texttt{model\_\allowbreak recommendations.json} file, which maps mission categories to empirically selected models.
    \item \texttt{code\_\allowbreak utils.py} extracts executable Python code from LLM responses and performs syntax validation prior to execution.
    \item \texttt{token\_\allowbreak utils.py} measures token usage for both local and cloud-based inference, enabling resource tracking and comparative analysis.
    \item \texttt{ardupilot\_\allowbreak wrapper.py} encapsulates MAVLink communication through the \texttt{ArduPilotWrapper} class, providing high-level functions for mission execution (\texttt{go\_\allowbreak to\_\allowbreak real\_\allowbreak world\_\allowbreak coords}, \texttt{move\_\allowbreak relative}, \texttt{start\_\allowbreak mission}) and telemetry tracking.
    \item \texttt{ardupilot\_\allowbreak utils.py} provides supporting utilities, including coordinate transformations, movement monitoring (\texttt{monitor\_\allowbreak movement\_\allowbreak completion}), stability detection, OpenStreetMap resolution, and MAVLink handling.
\end{itemize}

\subsection{Data Storage and Logging}

AERIAL maintains the following structured logs across both phases:

\begin{itemize}
    \item \textbf{Chat History:} Each session is stored in JSON
    format (\texttt{chat\_\allowbreak history/\allowbreak YYYY-MM-DD\_\allowbreak HH-MM-SS.json}),
    including prompts, model responses, and associated metrics.

    \item \textbf{Performance Metrics:} Aggregated benchmarking
    results are stored in CSV format
    (\texttt{datasets/\allowbreak combined\_\allowbreak models\_\allowbreak metrics.csv}), including
    mission category, model, latency, token usage, prompt count,
    and success statistics.

    \item \textbf{Model Recommendations:} The scenario-model
    mapping used during execution is stored in
    \texttt{model\_\allowbreak recommendations.json} and loaded at startup.
\end{itemize}

\section{Evaluation Methodology}\label{EM}

The evaluation followed the AERIAL two-phase architecture. During the \textit{Benchmarking Phase}, multiple LLMs were benchmarked across representative UAV mission scenarios to collect performance metrics, including response latency, token usage, and mission success rate. During the \textit{Execution Phase}, AERIAL applied the resulting scenario-model mapping to perform mission-aware model selection.

Both locally deployed and cloud-hosted models were evaluated to characterize trade-offs between inference speed, reasoning capability, and deployment flexibility. The evaluation focused on three aspects: (i) model performance across structured UAV control missions; (ii) the impact of mission-aware model selection; and (iii) the practical implications of hybrid edge-cloud operation.

\subsection{LLMs Under Evaluation}

Three locally deployed open-source LLMs were benchmarked to maintain continuity with prior work ~\cite{fluc2025} and one cloud-hosted model was included to evaluate hybrid deployment. The selected models represent different trade-offs between reasoning capability and inference efficiency:

\begin{itemize}
    \item \textbf{Qwen 2.5 Coder (14B):} a code-oriented model optimized for structured program synthesis and multi-step reasoning ~\cite{qwen2.5}.

    \item \textbf{Gemma 2 (9B):} a general-purpose decoder-only model offering a balance between reasoning quality and response latency ~\cite{gemma2}.

    \item \textbf{LLaMA 3.2 (3B):} a lightweight model designed for responsiveness and resource efficiency ~\cite{llama3.2}.

    \item \textbf{Gemini 2.5 Flash (Cloud):} a cloud-hosted model accessed via its official API, included to assess extended-context reasoning within the hybrid architecture ~\cite{gemini25}.
\end{itemize}

The inclusion of lightweight models aligns with recent work demonstrating effective UAV control using resource-constrained LLMs~\cite{LLMLand2025}, while the cloud-based model enables evaluation of hybrid edge-cloud inference strategies \cite{Hao2024}.

\subsection{Benchmarking Pipeline}

Each model-scenario pair was evaluated through 30 automated runs in ArduPilot SITL. The pipeline included prompt generation, mission code execution, telemetry monitoring, and scenario-specific validation. Performance metrics recorded for each run included LLM response latency, token usage, and mission success rate (see evaluation metrics defined in Section~\ref{EM:EM}).

The automated pipeline enabled consistent and repeatable evaluation across all configurations. The emphasis on structured benchmarking followed methodologies adopted in related code generation works~\cite{simcode2025}, where empirical comparison across models revealed mission-dependent performance differences.

All experiments were conducted on a system running Ubuntu 24.04.1 LTS (x86\_64), equipped with an AMD Ryzen 7 5800HS processor (8 cores / 16 threads, up to 4.46~GHz), 16~GB DDR4 RAM, an NVIDIA GeForce GTX 1650 Mobile (Max-Q) GPU with 4~GB VRAM, and NVMe SSD storage.

\subsection{Mission Scenarios and Validation Criteria} \label{subsec:scenarios}

Six representative mission categories were defined to assess different aspects of natural language interpretation and control logic generation. Validation thresholds were selected based on typical GPS accuracy (5-15 m horizontal)~\cite{Grayson2018}, barometric precision (1-3 m vertical)~\cite{Alberi2017}, and collision avoidance safety margins~\cite{Banerjee2020}. The validation for each scenario was performed automatically using telemetry logs collected during simulation.

\subsubsection{Mission Scenario: Coordinates}
\textbf{Scenario:} The model received explicit GPS coordinates (e.g., ``Go to 41.1783107 -8.591609 17'') and generated corresponding navigation commands. \textbf{Validation:} Success required the final UAV position to be within 15~m horizontally and 2~m vertically of the target. The geodesic distance (the shortest path between two points along a curved surface) was computed using the last telemetry reading.

\subsubsection{Mission Scenario: Location}
\textbf{Scenario:} The model resolved a named place (e.g., ``Go to FEUP'') via OpenStreetMap~\cite{openstreetmap} and generated navigation logic. \textbf{Validation:} The same criteria as the Coordinates scenario were applied, requiring 15 m horizontal and 2 m vertical accuracy.

\subsubsection{Mission Scenario: Diagonal}

\textbf{Scenario:} The instruction specified relative movement (e.g., ``Fly in a straight diagonal line for 200 meters''), testing dynamic waypoint computation. \textbf{Validation:} Distance traveled had to fall within 50\% of the requested value, and the ratio of latitude-to-longitude displacement had to remain within 20\% of unity to confirm diagonal movement.

\subsubsection{Mission Scenario: Waypoints}
\textbf{Scenario:} The model received five 3D waypoints, generating a traversal sequence through them. \textbf{Validation:} At least 90\% of waypoints had to be visited within a 15 m radius. Visitation was determined from telemetry logs.

\subsubsection{Mission Scenario: Obstacle}
\textbf{Scenario:} The model received an obstacle's dimensions and a destination, and generated trajectory logic to avoid it while reaching the target. \textbf{Validation:} The UAV had to travel at least 50\% of the target distance, maintain a minimum 2 m clearance from the obstacle volume, and avoid collision. All trajectory points were checked against reconstructed obstacle geometry.

\subsubsection{Mission Scenario: Algorithm}
\textbf{Scenario:} The model received parameters from natural language and invoked the \texttt{SUPPLY} optimization function using them. \textbf{Validation:} Two-stage validation was applied. First, the parameters passed to \texttt{SUPPLY} were verified by recomputing reference waypoints and comparing them against generated outputs, requiring at least 95\% agreement within horizontal tolerance. Second, waypoint traversal was validated by confirming that at least 90\% of generated waypoints were visited within 15~m. Both conditions had to be satisfied for success.

Mission success rate across all scenarios was determined automatically using validation routines implemented in the \texttt{TestValidator} class.

\subsection{Evaluation Metrics}
\label{EM:EM}

Each model was evaluated using three quantitative metrics:

\begin{itemize}
    \item \textbf{LLM response latency:}
    Time measured from the moment the HTTP request was sent to the LLM endpoint (local Ollama or cloud Gemini API) until the complete response was received. This measurement includes model inference time, API latency, and response transmission, but excludes code parsing, validation, and UAV mission execution in SITL. The metric therefore reflects LLM code generation latency only.

    \item \textbf{Token usage:}
    Total number of tokens processed by the LLM, including both input tokens (system prompt, mission context, and user command) and output tokens (generated Python code and accompanying text). Token count was used as an indicator of computational cost for local inference and billing cost for cloud-based models.

    \item \textbf{Success rate:}
    Percentage of test runs in which the generated code executed without runtime errors and the UAV mission satisfied the scenario-specific validation criteria defined in
    Section~\ref{subsec:scenarios}. Success rate was computed using Eq.~(\ref{eq:success_rate}). Mission outcomes were determined automatically using the
    \texttt{TestValidator} class.
\end{itemize}

\begin{equation}
\label{eq:success_rate}
\text{Success rate} = \frac{N_{\text{success}}}{N_{\text{total}}} \times 100\%
\end{equation}

In Eq.~(\ref{eq:success_rate}), $N_{\text{success}}$ indicates the number of successful missions and $N_{\text{total}}$ indicates the total number of attempts.

These metrics reflect the computational cost of LLM inference exclusively. UAV propulsion energy and onboard power consumption are governed by flight dynamics and hardware independently of which LLM generates the mission code, and are outside the scope of this work.

\section{Evaluation Results}\label{ER}

This section presents the results obtained from AERIAL's two-phase evaluation.

\subsection{Benchmarking Results}

Table~\ref{tab:performance_results} summarizes the benchmarking results across all evaluated models and mission scenarios, according to the metrics defined in Section~\ref{EM}. The results reflect comparative performance trends observed under SITL-based simulation. Differences in success rates and LLM response latency are interpreted in the context of the scenario-specific validation criteria described in Section~\ref{subsec:scenarios}.

\begin{table*}[t]
\centering
\caption{Benchmarking results across models and mission scenarios. Values represent mean $\pm$ standard deviation over 30 runs per model-scenario pair.}
\label{tab:performance_results}
\begin{tabular}{llccc}
\toprule
\textbf{Scenario} & \textbf{Model} & \textbf{Response Latency (s)} & \textbf{Token Count} & \textbf{Success Rate (\%)} \\
\midrule

\multirow{4}{*}{Coordinates}
& LLaMA 3.2 & $4.9 \pm 1.0$ & $58.2 \pm 16.3$ & 53 \\
& \textbf{Gemma 2} & $\mathbf{12.9 \pm 0.3}$ & $\mathbf{45.6 \pm 0.2}$ & \textbf{100} \\
& Qwen 2.5 Coder & $21.0 \pm 2.1$ & $56.9 \pm 6.9$ & 53 \\
& Gemini 2.5 Flash & $6.9 \pm 3.8$ & $38.1 \pm 1.4$ & 90 \\
\midrule

\multirow{4}{*}{Location}
& LLaMA 3.2 & $3.6 \pm 0.3$ & $42.7 \pm 3.8$ & 0 \\
& \textbf{Gemma 2} & $\mathbf{6.9 \pm 0.2}$ & $\mathbf{21.9 \pm 0.3}$ & \textbf{93} \\
& Qwen 2.5 Coder & $17.2 \pm 0.9$ & $45.0 \pm 2.5$ & 0 \\
& Gemini 2.5 Flash & $8.3 \pm 4.9$ & $26.0 \pm 6.3$ & 50 \\
\midrule

\multirow{4}{*}{Diagonal}
& LLaMA 3.2 & $9.6 \pm 3.1$ & $116.2 \pm 35.6$ & 10 \\
& \textbf{Gemma 2} & $\mathbf{9.4 \pm 0.2}$ & $\mathbf{31.8 \pm 0.6}$ & \textbf{100} \\
& Qwen 2.5 Coder & $23.8 \pm 5.1$ & $65.7 \pm 15.7$ & 7 \\
& Gemini 2.5 Flash & $9.0 \pm 2.5$ & $37.3 \pm 6.6$ & 87 \\
\midrule

\multirow{4}{*}{Waypoints}
& LLaMA 3.2 & $31.2 \pm 5.6$ & $337.5 \pm 57.1$ & 3 \\
& Gemma 2 & $27.9 \pm 7.6$ & $91.9 \pm 29.4$ & 0 \\
& \textbf{Qwen 2.5 Coder} & $\mathbf{114.1 \pm 16.2}$ & $\mathbf{298.5 \pm 43.5}$ & \textbf{37} \\
& Gemini 2.5 Flash & $13.3 \pm 5.4$ & $58.4 \pm 55.2$ & 23 \\
\midrule

\multirow{4}{*}{Obstacle}
& LLaMA 3.2 & $13.8 \pm 4.8$ & $154.5 \pm 50.4$ & 10 \\
& Gemma 2 & $23.7 \pm 4.2$ & $79.5 \pm 15.9$ & 3 \\
& \textbf{Qwen 2.5 Coder} & $\mathbf{50.4 \pm 8.0}$ & $\mathbf{140.7 \pm 23.5}$ & \textbf{57} \\
& Gemini 2.5 Flash & $9.3 \pm 1.1$ & $78.4 \pm 8.7$ & 40 \\
\midrule

\multirow{4}{*}{Algorithm}
& LLaMA 3.2 & $4.7 \pm 0.7$ & $55.6 \pm 8.7$ & 53 \\
& Gemma 2 & $16.3 \pm 1.1$ & $56.8 \pm 4.0$ & 67 \\
& Qwen 2.5 Coder & $22.9 \pm 1.5$ & $59.5 \pm 4.2$ & 70 \\
& \textbf{Gemini 2.5 Flash} & $\mathbf{2.1 \pm 0.4}$ & $\mathbf{51.7 \pm 2.7}$ & \textbf{70} \\
\bottomrule
\end{tabular}
\end{table*}

\subsubsection{LLM Response Latency Analysis}
LLM response latency presented distinct patterns across models and mission scenarios. LLaMA 3.2 achieved the lowest latency for simple navigation tasks, requiring 3.6~s for location-based missions and 4.9~s for coordinate navigation. Latency increased for more complex scenarios, with waypoint sequencing requiring 31.2 s.

Gemma 2 maintained relatively stable latency across navigation tasks (6.9-12.9 s), with a moderate increase for waypoint sequencing (27.9 s). Qwen 2.5 Coder showed consistently higher latency, ranging from 17.2~s for location tasks to 114.1~s for waypoint sequencing.

The cloud-based Gemini 2.5 Flash achieved competitive latency across most scenarios (2.1-13.3 s), including the lowest latency for algorithm invocation (2.1 s). These results highlight the trade-off between model size, reasoning complexity, and inference time.

\subsubsection{Token Efficiency Analysis}
Token consumption reflected differences in output verbosity and code structure. Gemma 2 generated comparatively concise outputs, with token counts between 21.9 and 91.9 across scenarios. Gemini 2.5 Flash showed similar efficiency (26.0-78.4 tokens), although waypoint sequencing exhibited higher variance (58.4 $\pm$ 55.2 tokens).

LLaMA 3.2 demonstrated moderate token usage for simpler tasks (42.7-116.2 tokens), but token counts increased substantially for waypoint sequencing (337.5 tokens). Qwen 2.5 Coder produced more verbose outputs overall, particularly in waypoint (298.5 tokens) and obstacle (140.7 tokens) scenarios, consistent with its structured reasoning style.

\subsubsection{Success Rate Analysis}
Mission success rates showed clear scenario-dependent behavior. Gemma 2 achieved high success rates (93-100\%) in coordinate, location, and diagonal scenarios, but performance decreased in waypoint (0\%) and obstacle (3\%) scenarios, indicating difficulty with multi-step planning and constraint reasoning.

Qwen 2.5 Coder achieved the highest success rate in obstacle scenarios (57\%) and moderate performance in waypoint sequencing (37\%), while success in diagonal movement remained limited (7\%).

LLaMA 3.2 showed limited reliability across scenarios, with success rates not exceeding 53\% and no successful runs in location-based tasks. Gemini 2.5 Flash demonstrated more balanced performance (40-90\%) across most scenarios, including strong results in coordinate navigation (90\%) and diagonal movement (87\%), although waypoint sequencing remained challenging (23\%).

\subsection{Performance Characteristics}

The evaluation results revealed significant model-scenario dependencies that validate AERIAL's adaptive selection approach.

\subsubsection{Model-specific Observations}

This section presents the individual performance of each LLM across the benchmarking suite.
\textit{LLaMA 3.2} achieved the fastest response times (3.6-9.6 seconds) but demonstrated insufficient reasoning capability, with success rates not exceeding 53\% and complete failure in location-based tasks (0\%). The model's 3B parameter count limits its ability to handle complex spatial reasoning, making it unsuitable for autonomous mission-critical deployments despite its speed advantage.

\textit{Gemma 2} delivered near-perfect success rates (93-100\%) for basic navigation primitives with the most efficient token usage (21.9-91.9 tokens). However, it completely fails in waypoint sequencing (0\%) and obstacle avoidance (3\%), revealing fundamental limitations in multi-step planning and conditional logic that restrict its applicability to straightforward navigation tasks.

\textit{Qwen 2.5 Coder} was the best among all the models in constraint reasoning scenarios, achieving 57\% success in obstacle avoidance, nearly 20$\times$ better than Gemma 2. Its code-specialized training enables structured reasoning about spatial constraints, justifying its deployment despite increased latency (17.2-114.1 seconds) and verbose outputs (140.7-298.5 tokens) for tasks requiring reasoning depth over response speed.

\textit{Gemini 2.5 Flash} demonstrated balanced performance across scenarios (40-90\% success) with competitive latency (2.1-13.3 seconds) and efficient token usage. Its strength in coordinate navigation (90\%) and exceptionally fast algorithm invocation (2.1 seconds, 70\% success) positions it as a viable cloud-based alternative when network connectivity permits.

\subsubsection{Scenario-specific Observations}

This analysis examined the influence of mission type on success rates and identified common failure patterns across tasks.

Waypoint sequencing was the most challenging scenario, with no locally deployed model exceeding 37\% success. The cloud-based Gemini 2.5 Flash showed similarly limited performance. These results indicate that multi-step spatial sequencing remains difficult across the evaluated models.

Obstacle avoidance exhibited the widest performance range (3-57\%), reflecting strong sensitivity to model selection in constraint-driven scenarios.

Analysis of 397 failed runs across six scenarios revealed mission-aware error patterns. Location navigation showed a high proportion of name-related errors (77.9\%), frequently caused by references to undefined variables (53 instances), indicating difficulties in strict API adherence. Obstacle avoidance produced 87 failures, primarily validation errors (74.7\%), showing that models often generated syntactically valid but geometrically incorrect trajectories.

Waypoint sequencing produced 101 failures with a diverse error distribution. Models frequently attempted to define helper functions not available in the provided control interface, indicating deviation from the specified API structure. LLaMA 3.2 showed recurring issues in mathematical operations, including missing references to the \texttt{math} module (6 instances) and undefined variable errors (9 instances).

Overall, performance varied substantially across mission types and no single model consistently achieved high reliability in all scenarios. The observed model-task dependencies support the use of mission-aware selection to align model characteristics with scenario requirements.

\subsection{Mission-aware Model Selection Results}

Following benchmarking, performance metrics were aggregated to construct a scenario-model selection policy. The system generated a JSON-based mapping associating each mission category with the model achieving the highest empirical success rate while considering latency and token usage. The mapping was computed using a weighted scoring function over success rate (0.65), average response latency (0.20), and average token count (0.15) with ties broken in favor of the higher success rate. The weights are configurable, therefore operators can adjust the weighting to reflect their deployment-specific priorities. Success rate is used directly, without normalization while response latency and token count are min-max normalized across the models competing within each mission category, and the result is then inverted so that lower value yields a higher score. The resulting mapping was derived exclusively from benchmarking data and remained fixed during execution.

The benchmark-driven policy enabled mission-aware routing, allowing each mission type to be processed by the model that demonstrated comparatively stronger performance during benchmarking (e.g., Qwen 2.5 Coder for obstacle scenarios and Gemma 2 for basic navigation tasks). The mapping also supported selection of lightweight models to routine missions while reserving larger models for scenarios requiring structured reasoning. Because the mapping was data-driven, it could be recomputed periodically as additional benchmarking results became available.

\subsubsection{Learned Model-Selection Mapping}

The final model-selection mapping, stored in \texttt{model\_\allowbreak recommendations.json}, was:

\begin{verbatim}
{
    "coordinates": "gemma2:9b",
    "location": "gemma2:9b",
    "diagonal": "gemma2:9b",
    "waypoints": "qwen2.5-coder:14b",
    "obstacle": "qwen2.5-coder:14b",
    "algorithm": "qwen2.5-coder:14b"
}
\end{verbatim}

This mapping reflected the empirically strongest-performing model for each mission category under the evaluated conditions, considering success rate together with latency and token usage. It represents a snapshot corresponding to the specific models and scenarios included in this study.

\subsubsection{Benefits of Adaptive Selection}

Compared with static model configurations, the adaptive selection mechanism exhibited several practical advantages.

\begin{enumerate}
    \item \textbf{Mission-aware Reliability:}
    Missions were routed to models that demonstrated higher
    empirical success rates within the corresponding scenario
    category. For example, obstacle avoidance tasks were
    assigned to Qwen 2.5 Coder, which achieved higher success
    rates than alternative models in that scenario.

    \item \textbf{Resource Allocation Efficiency:}
    Larger models were reserved for scenarios requiring structured reasoning, while lighter models were assigned
    to navigation tasks where they achieved comparable or
    superior reliability.

    \item \textbf{Data-Driven Policy Formation:}
    The selection mapping was derived directly from measured
    performance rather than manual configuration, reducing
    subjective model assignment.

    \item \textbf{Updateability:}
    Because the mapping was stored externally in JSON format,
    it could be recomputed when new benchmarking data or models
    were introduced, without modifying system logic.

    \item \textbf{Hybrid Deployment Flexibility:}
    Although Gemini 2.5 Flash achieved competitive performance
    in several scenarios, the default mapping prioritized local
    models to preserve offline capability. Cloud-based models
    remained available as optional alternatives when network
    connectivity permitted.
\end{enumerate}

\subsubsection{Execution Phase Example}

\begin{figure}[t]
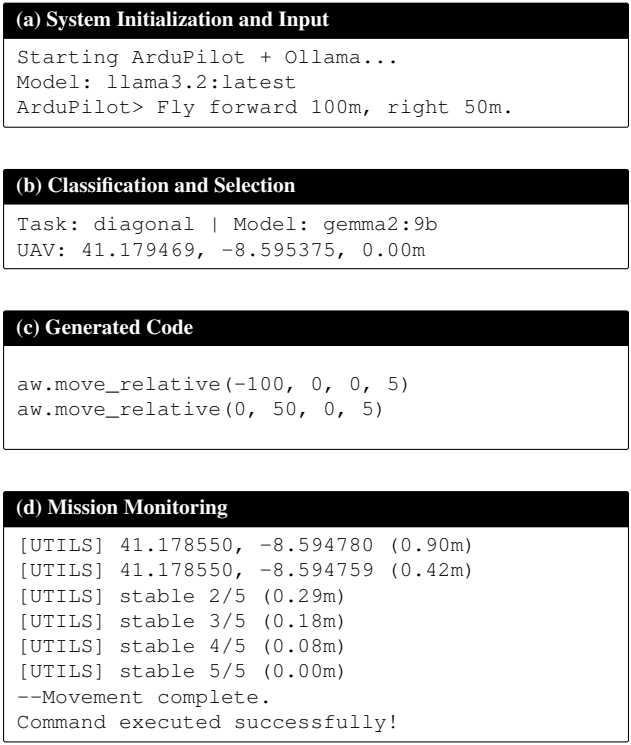

\centering
\footnotesize
\begin{tcolorbox}[colback=white,colframe=black,boxrule=0.4pt,
arc=0.5pt,left=2pt,right=2pt,top=1pt,bottom=1pt,
title={\footnotesize\textbf{(a) System Initialization and Input}}]
\begin{verbatim}
Starting ArduPilot + Ollama...
Model: llama3.2:latest
ArduPilot> Fly forward 100m, right 50m.
\end{verbatim}
\end{tcolorbox}

\medskip

\begin{tcolorbox}[colback=white,colframe=black,boxrule=0.4pt,
arc=0.5pt,left=2pt,right=2pt,top=1pt,bottom=1pt,
title={\footnotesize\textbf{(b) Classification and Selection}}]
\begin{verbatim}
Task: diagonal | Model: gemma2:9b
UAV: 41.179469, -8.595375, 0.00m
\end{verbatim}
\end{tcolorbox}

\medskip

\begin{tcolorbox}[colback=white,colframe=black,boxrule=0.4pt,
arc=0.5pt,left=2pt,right=2pt,top=1pt,bottom=1pt,
title={\footnotesize\textbf{(c) Generated Code}}]
\begin{lstlisting}[language=Python,basicstyle=\footnotesize\ttfamily]
aw.move_relative(-100, 0, 0, 5)
aw.move_relative(0, 50, 0, 5)
\end{lstlisting}
\end{tcolorbox}

\medskip

\begin{tcolorbox}[colback=white,colframe=black,boxrule=0.4pt,
arc=0.5pt,left=2pt,right=2pt,top=1pt,bottom=1pt,
title={\footnotesize\textbf{(d) Mission Monitoring}}]
\begin{verbatim}
[UTILS] 41.178550, -8.594780 (0.90m)
[UTILS] 41.178550, -8.594759 (0.42m)
[UTILS] stable 2/5 (0.29m)
[UTILS] stable 3/5 (0.18m)
[UTILS] stable 4/5 (0.08m)
[UTILS] stable 5/5 (0.00m)
--Movement complete.
Command executed successfully!
\end{verbatim}
\end{tcolorbox}

\caption{Execution phase workflow illustrating (a) system initialization and user input, (b) task classification and model selection, (c) generated control code, and (d) runtime monitoring with stability detection.}
\label{fig:execution_example}

\end{figure}

Fig.~\ref{fig:execution_example} illustrates the execution workflow for a diagonal movement mission combining forward and lateral navigation. The user provided the natural language command \textit{``Fly forward for 100 meters, then go to the right for 50 meters''} through the interactive interface (Panel~(a)).

\textbf{Classification and Selection:} The embedded LLaMA 3.2 model analyzed the prompt and classified it as a \texttt{diagonal} scenario based on the sequential relative movements (Panel~(b)). The system reported the current UAV position (Latitude 41.179469, Longitude -8.595375, Altitude 0.00~m) and queried the Model Selection Engine, which loaded the stored \texttt{model\_\allowbreak recommendations.json} mapping. According to the learned mapping, the system selected \texttt{gemma2:9b} for this task category.

\textbf{Code Generation:} The selected model generated Python code through the Ollama interface using the ArduPilot wrapper library's \texttt{move\_\allowbreak relative} function (Panel~(c)). The code issued two sequential commands: a 100-meter forward movement (North direction, represented as -100 in NED coordinates) followed by a 50-meter rightward movement (East direction). Each command included a 5-second timeout parameter for monitoring.

\textbf{Mission Execution:} The Runtime Supervision Layer monitored UAV telemetry throughout the mission (Panel~(d)), providing continuous position updates. The system tracked movement progression and applied the stability detection mechanism across five consecutive readings. Recorded position changes decreased from 0.42~m to 0.00~m, satisfying the stability threshold and triggering confirmation of movement completion.

\begin{figure}[t]
\centering
\includegraphics[width=\columnwidth]{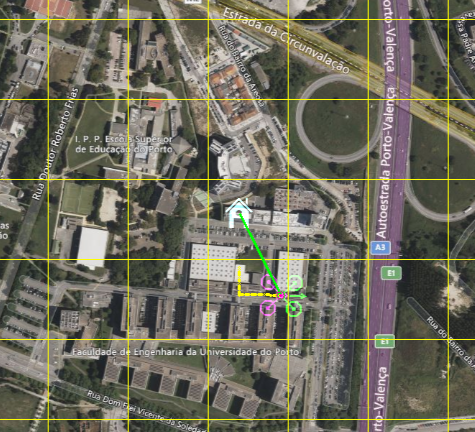}
\caption{ArduPilot map visualization showing the executed flight path. The yellow
dotted line represents the actual trajectory followed by the UAV, while the green
diagonal line indicates the direct path from the initial position to the final
destination.}
\label{fig:trajectory_map}
\end{figure}

\textbf{Trajectory Visualization:} Fig.~\ref{fig:trajectory_map} shows the ArduPilot map visualization of the executed trajectory. The yellow dotted line represents the recorded UAV path, while the green line indicates the direct displacement between the initial and final positions. The visualization confirms sequential execution of the forward and lateral movements.

\textbf{Outcome:} The mission concluded with a final recorded position of (41.178550, -8.594753). The system reported successful command execution following stability confirmation. This example illustrates the interaction between task classification, model selection, code generation, and runtime supervision during execution.

\subsection{Summary}

The evaluation assessed AERIAL through systematic benchmarking of four LLMs across six mission scenarios. The results highlighted clear model-scenario dependencies, supporting the design of the two-phase architecture. Automated benchmarking identified distinct performance patterns, including high reliability of Gemma 2 in basic navigation tasks and stronger constraint-handling performance of Qwen 2.5 Coder in obstacle scenarios.

In the execution phase, the learned mapping enabled mission-aware routing, and the task classifier achieved 100\% accuracy on the structured benchmarking prompts used during system testing. The six prompts were designed as direct instances of the six predefined mission categories and to be linguistically distinct in structure, which contributes to reliable classification within this distribution. Overall, the findings indicate that benchmark-driven model selection provides more consistent reliability and improved resource allocation compared to static single-model configurations under the evaluated conditions.

\section{Case Study: Validation with the SUPPLY Algorithm}\label{CS}

To evaluate AERIAL's ability to interface with domain-specific optimization algorithms in FN scenarios, this section presents a case study integrating AERIAL with the SUPPLY algorithm~\cite{Ribeiro2024}. The objective was to assess whether the mission-aware model-selection mechanism maintained reliable function invocation when processing structured, parameter-dependent mission specifications.

\subsection{SUPPLY Algorithm Overview}

SUPPLY is an energy-aware UAV positioning algorithm designed for FNs. It computes 3D UAV trajectories that minimize propulsion energy while satisfying QoS requirements for Ground Users (GUs)~\cite{Ribeiro2024}. In this study, SUPPLY was treated as a black-box optimization module. Its internal convergence behavior and optimality properties were outside the scope of the present evaluation. The integration focused exclusively on verifying correct parameter extraction, function invocation, and mission execution through AERIAL's control pipeline.

\subsection{Test Design}

The evaluation included 50 distinct mission configurations, each varying the spatial coordinates (\texttt{x}, \texttt{y}, \texttt{z}) and traffic demands (\texttt{traffic}) of two GUs. Each configuration was executed once per model (LLaMA 3.2, Gemma 2, Qwen 2.5 Coder, and Gemini 2.5 Flash), resulting in 200 total test runs.

Natural language prompts followed the structure:

\textit{``Upload and start the mission for a UAV with 2 GUs, whose x are [x1,x2], y are [y1,y2], z are [z1,z2], and traffic [t1,t2].''}

Each LLM was required to extract the parameters from the prompt and generate a valid function call invoking the SUPPLY algorithm. A test run was classified as successful when the generated code correctly parsed the parameters, invoked the SUPPLY function with appropriate syntax, produced valid waypoints, and completed mission execution without runtime errors.

\subsection{Results}

Table~\ref{tab:supply_results} summarizes the performance metrics obtained for the SUPPLY case study across the evaluated models.

\begin{table}[t]
\centering
\caption{Performance metrics for SUPPLY integration (mean $\pm$ std).}
\label{tab:supply_results}
\setlength{\tabcolsep}{4pt}
\small
\begin{tabular}{lccc}
\toprule
\textbf{Model} & \textbf{Time (s)} & \textbf{Tokens} & \textbf{Succ. (\%)} \\
\midrule
LLaMA 3.2 & $6.58 \pm 2.27$ & $99.3 \pm 50.6$ & 44 \\
Gemma 2 & $16.74 \pm 5.86$ & $75.3 \pm 39.3$ & 66 \\
\textbf{Qwen 2.5 Coder} & $\mathbf{23.63 \pm 6.25}$ & $\mathbf{79.1 \pm 21.8}$ & \textbf{70} \\
\textbf{Gemini 2.5 Flash} & $\mathbf{4.82 \pm 7.26}$ & $\mathbf{87.8 \pm 12.1}$ & \textbf{70} \\
\bottomrule
\end{tabular}
\end{table}

Qwen 2.5 Coder and Gemini 2.5 Flash achieved the highest success rates (70\%), while Gemma 2 reached 66\%, generating correct and executable function calls in the majority of configurations. LLaMA 3.2 achieved 44\% success, with failures primarily associated with incorrect parameter formatting or incomplete function invocation syntax.

LLaMA 3.2 exhibited the lowest response latency among local models (6.58~s on average) but the highest token usage (99.3 tokens), likely reflecting output inflation from
failed or incomplete invocations. This efficiency corresponded to reduced reliability in structured function invocation. Gemma 2 showed higher latency (16.74~s) with more compact token usage (75.3 tokens) and achieved 66\% success. Qwen 2.5 Coder presented higher latency (23.63~s) and moderate token counts (79.1 tokens), while achieving 70\% success. Gemini 2.5 Flash achieved comparable success (70\%) with the lowest overall latency (4.82~s), though token usage was higher than local models at comparable success rates (87.8 tokens).

Across the 75 failed runs, model-specific error patterns were observed. LLaMA 3.2 accounted for 28 failures, primarily due to generic exceptions and name errors, including incorrect references to \texttt{python} as a variable. Gemma 2 produced 17 failures, largely associated with incorrect argument passing or attempts to access SUPPLY outputs prior to invocation. Gemini 2.5 Flash and Qwen 2.5 Coder each produced 15 failures, with name errors and generic execution exceptions as the dominant categories.

These results suggest that structured parameter extraction and function invocation remain challenging across evaluated LLMs, particularly under strict API constraints.

\subsection{Summary}

The SUPPLY case study evaluated AERIAL's integration with a domain-specific optimization algorithm through natural language interfaces. The results showed that the code-oriented model (Qwen 2.5 Coder) and the cloud-based model (Gemini 2.5 Flash) achieved the highest reliability (70\%) in structured function invocation tasks, while the general-purpose Gemma 2 achieved 66\% success. LLaMA 3.2 exhibited lower reliability (44\%) despite being the fastest among locally deployed models.

These findings were consistent with the broader benchmarking results presented in Section~\ref{ER}, where structured and constraint-driven tasks benefited from models with stronger reasoning capabilities. The case study indicated that the mission-aware model-selection mechanism routed algorithm-invocation tasks to comparatively reliable models under the evaluated conditions. Overall, the integration results suggest that the framework can support parameter-dependent optimization workflows in addition to basic navigation primitives.

\section{Discussion}\label{D}

This section discusses the broader implications of AERIAL's design and evaluation results within the context of LLM-based autonomous systems.

\subsection{Key Findings}

The evaluation across six mission scenarios and four LLMs revealed consistent mission-aware performance variation. The results supported the architectural premise of separating benchmarking from execution through a two-phase framework.

First, the automated benchmarking pipeline enabled systematic and reproducible benchmarking at a scale that may be difficult to achieve through manual evaluation. Thirty iterations per configuration (720 benchmarking runs) enabled estimation of confidence intervals and reduced variance in performance estimates. Automated validation functions ensured consistent success criteria.

Second, benchmark-driven model selection identified distinct model-task affinities. Navigation-oriented scenarios were most reliably handled by Gemma 2, while constraint-driven tasks such as obstacle avoidance showed comparatively higher success rates with Qwen 2.5 Coder. These results suggest that model reliability in specific operational contexts may be more consequential than general reasoning capacity. The observed specialization patterns support mission-aware model selection rather than uniform deployment of a single model.

Third, the SUPPLY case study extended the evaluation to structured, parameter-dependent function invocation. Across 200 runs, Qwen 2.5 Coder and Gemini 2.5 Flash achieved 70\% success, while Gemma 2 achieved 66\%. These results indicate that LLM-based control layers can interface with domain-specific optimization modules under defined API constraints, although reliability remains below deterministic software baselines.

The observed model-task dependency patterns align with trends presented in related benchmarking works. For example, the SIMCODE benchmark~\cite{simcode2025} concluded mission-aware performance variation across LLMs in network simulation code generation. This consistency indicates that systematic benchmarking may provide value in other autonomous systems that require translation from natural language to executable control logic.

\subsection{Practical Implications for FNs}

From an operational perspective, AERIAL reduces manual model configuration by implementing selection into a data-driven mapping. The modular architecture ensures compatibility with standard UAV control stacks (e.g., MAVLink and ArduPilot) while enabling independent upgrades of LLM components. Hybrid edge-cloud compatibility supports deployment under variable connectivity conditions. Local models preserve offline autonomy, while cloud models remain available when network access permits. This flexibility may be relevant in FN deployments characterized by limited Internet connectivity. It is worth noting that AERIAL is not proposed as an alternative to rule-based systems, reinforcement learning planners, or traditional mission planners. It operates as a natural language interface above such systems.

\subsection{Limitations}
Several limitations should be considered.
The evaluation was conducted exclusively in simulation using ArduPilot SITL; real-world deployment may introduce additional uncertainties related to actuation delays and environmental variability.

Performance results were obtained on a specific hardware configuration and may vary across computational platforms.

Furthermore, the evaluated scenarios covered six mission categories and may not represent the full operational spectrum of FN deployments. In addition, the evaluation was limited to single-UAV missions. AERIAL's mission-aware selection and hybrid execution mechanisms were not evaluated in coordinated multi-UAV scenarios. The adaptive selection mechanism relied on precomputed scenario-model mappings generated during benchmarking, and the mapping was not evaluated under a cross-validation protocol or validated on held-out prompts. The execution phase results therefore reflect operational behavior within the prompts used for benchmarking rather than unseen inputs. Periodic re-evaluation may be required to incorporate new models or evolving performance characteristics. The current implementation did not include automated remapping triggers.

The hybrid edge-cloud architecture was evaluated in terms of functional integration and LLM-level performance, through mission success rate, response latency, and token usage. It was not evaluated under degraded connectivity, simulated cloud outages, handover and fallback behavior, or varying communication overhead; these aspects are directions for future work.

The performance levels observed in complex mission types represent a further limitation. In addition, AERIAL is a benchmark-driven model-selection framework and is not designed to function as a safety-control mechanism; it does not replace deterministic collision-avoidance or mission-planning modules required for safe operational deployment.
Improving performance on these mission types through prompt engineering, or integration of more capable models, is a necessary step before AERIAL can be considered for deployment beyond research contexts.

\section{Conclusions}\label{C}

This paper presented AERIAL, a benchmark-driven framework for natural language-driven UAV control in Flying Networks that combines systematic model benchmarking, adaptive LLM selection, and hybrid edge-cloud execution. By decoupling model evaluation from mission execution through a two-phase design, AERIAL enables data-driven model selection based on empirical performance rather than fixed manual configuration. The evaluation results demonstrate that LLM performance varies significantly across mission types, and that exploiting these differences through adaptive selection improves the reliability and operational robustness of autonomous UAV control. Overall, AERIAL establishes a practical foundation for benchmark-driven, language-based UAV autonomy in Flying Networks.

Future work will focus on extending the framework beyond the current UAV simulation- and emulation-based settings, particularly through real-world UAV validation and support for more complex mission profiles, including coordinated multi-UAV scenarios.
Also, it will consider the extension of the benchmarking suite to a broader and more diverse set of mission categories, as well as the evaluation under realistic FN networking conditions. Additional complementary efforts will address online adaptation of model-selection policies and the integration of complementary reasoning and control mechanisms to further improve robustness in dynamic operating conditions.

{\small
\bibliographystyle{IEEEtran}
\bibliography{refs}
}

\end{document}